\documentclass[
 reprint,
 superscriptaddress,
 longbibliography,
 amsmath,amssymb,
 aps,pra,
]{revtex4-2}

\usepackage{graphicx}
\usepackage{amsthm}
\usepackage{hyperref}

\theoremstyle{plain}
\newtheorem{theorem}{Theorem}

\theoremstyle{definition}
\newtheorem{definition}{Definition}
\newtheorem{example}{Example}

\begin{document}

\title{Gradient Blindness in Adaptive Architecture Search for IQP Born Machines}

\author{Sebastian W\"olckert}
\email{Contact author: sebastian.woelckert@ifi.lmu.de}
\affiliation{LMU Munich, Munich, Germany}

\author{Gerhard Stenzel}
\affiliation{LMU Munich, Munich, Germany}

\author{Michael Poppel}
\affiliation{LMU Munich, Munich, Germany}
\affiliation{Aqarios GmbH, Munich, Germany}

\author{Claudia Linnhoff-Popien}
\affiliation{LMU Munich, Munich, Germany}

\author{Jonas Stein}
\affiliation{LMU Munich, Munich, Germany}

\date{\today}

\begin{abstract}
Instantaneous quantum polynomial (IQP) Born machines learn a probability distribution from data with a circuit of commuting gates that a classical computer trains at hundreds of qubits.
Adaptive algorithms grow a sparse circuit from a small seed by adding in every round the candidate gates with the largest loss gradient.
Such algorithms never add a candidate whose gradient vanishes, so the choice of seed decides which circuits they can reach.
Here we show that a candidate gate has a nonzero gradient for some angles of the circuit exactly when its generator is the sum of an odd number of linearly independent generators already in the circuit.
The common seed of one-qubit gates therefore never adds a gate on an even number of qubits, which exact gradients on 12 qubits confirm.
A seed of one-qubit gates plus the edges of the Chow-Liu tree sees every candidate, and its circuits match the full circuits on 12 and 16 qubits with 10 to 13 percent of their parameters.
At 484 qubits, a heredity pool keeps the search over higher-order gates feasible, and the circuits grown from this seed reach a lower test loss than the dense two-body circuit with 6 percent of its parameters, although the higher-order gates add no gain.
\end{abstract}

\maketitle

\section{Introduction}\label{sec:intro}

Generative models learn a probability distribution from data and draw new samples from it.
For binary data, the model has to assign a probability to each of the $2^n$ bitstrings of length $n$.
A quantum circuit Born machine (QCBM) represents this distribution by a parameterized $n$-qubit state and draws a bitstring by measuring the state in the computational basis \cite{liuDifferentiableLearningQuantum2018,benedettiGenerativeModelingApproach2019}.

Parameterized instantaneous quantum polynomial (IQP) circuits are QCBMs that consist of commuting gates, all diagonal in the Pauli $X$ basis \cite{coyleBornSupremacyQuantum2020}.
Sampling from their output distributions is hard for a classical computer under plausible complexity assumptions \cite{bremnerClassicalSimulationCommuting2011,bremnerAveragecaseComplexityApproximate2016,bremnerAchievingQuantumSupremacy2017,hangleiterComputationalAdvantageQuantum2023}, whereas training is not.
Recio-Armengol et al.~\cite{recio-armengolTrainClassicalDeploy2025} show that a classical computer trains these circuits efficiently at hundreds of qubits and hundreds of thousands of parameters while the quantum computer only performs the sampling.
Such training has also been shown to generalize well to unseen data for specific tasks \cite{baumannParitySupervisionDriver2026}.

If these circuits contain all gates on one and two qubits \cite{recio-armengolTrainClassicalDeploy2025}, the number of parameters grows quadratically with the number of qubits.
Sparse circuits are also hard to sample from \cite{bremnerAchievingQuantumSupremacy2017}, but finding one that fits the data requires knowledge of the structure of the problem \cite{bakoProbleminformedGraphicalQuantum2026}.
Other methods derive the structure from pairwise correlations of the data \cite{liuDifferentiableLearningQuantum2018}, which leaves out the interactions of higher order.

Adaptive algorithms grow a circuit from a small seed, the starting set of gates.
In every round, they score each candidate gate in a pool by the loss gradient with respect to its angle at zero, where the gate is still the identity, and add the gates with the largest scores.
The adaptive derivative-assembled pseudo-Trotter variational quantum eigensolver (ADAPT-VQE) introduced this principle in quantum chemistry \cite{grimsleyAdaptiveVariationalAlgorithm2019}, the adaptive quantum approximate optimization algorithm (ADAPT-QAOA) applies it to combinatorial optimization \cite{zhuAdaptiveQuantumApproximate2022}, and Li and Cheng~\cite{liAdaptiveCircuitLearning2023} apply it to Born machines.

These algorithms add the gates with the largest gradients and are therefore blind to every candidate whose gradient is zero.
We call such a candidate blind and every other candidate visible.
In ADAPT-VQE and ADAPT-QAOA, a symmetry of the Hamiltonian or of the state forces the gradient of some candidates to zero in every round, and the pool is designed without them \cite{tangQubitADAPTVQEAdaptiveAlgorithm2021,zhuAdaptiveQuantumApproximate2022}.
For IQP circuits, we find that the blindness arises from the structure of the circuit, through a symmetry that the seed itself carries.
The circuits of Li and Cheng~\cite{liAdaptiveCircuitLearning2023}, whose gates do not commute, do not show this blindness.

In this work, we derive an exact criterion for when a gate not yet in the circuit has a nonzero gradient.
The criterion builds on the expansion of the expectation values by Shen et al.~\cite{shenCharacterizingTrainabilityInstantaneous2026}, who study the gradients of gates already in the circuit.
We trace the blindness to the gates that are currently active in the circuit.
We prove that if the generators of these gates lie on an affine hyperplane over $\mathbb{F}_2$, the circuit never leaves this hyperplane under exact gradients, while otherwise no candidate in their span stays blind permanently.
With estimated gradients, the noise lets such a seed leave the hyperplane, but in our experiments its circuits end at a higher test loss.
We then scale the algorithm, which we call ADAPT-IQP, to $484$ qubits with a heredity pool, which holds only one-qubit extensions of the active gates and keeps the search over gates of higher order feasible, although on the D-Wave dataset these gates give no gain over two-body gates.

The paper is organized as follows.
Section~\ref{sec:prelims} introduces IQP circuits, the Kullback-Leibler (KL) divergence and the maximum mean discrepancy (MMD) as loss functions, and the Chow-Liu (CL) tree.
Section~\ref{sec:related} reviews related work.
Section~\ref{sec:theory} derives the visibility criterion and the theorems on affine active sets.
Section~\ref{sec:method} describes ADAPT-IQP, its seeds and the heredity pool, and Sec.~\ref{sec:setup} gives the numerical setup.
Section~\ref{sec:results} reports the results on Two Moons with $12$ qubits and exact gradients, on Binary Blobs with $16$ qubits and on the D-Wave dataset with $484$ qubits, followed by the ablations.
Section~\ref{sec:conclusion} summarizes the results and their limits.

\section{Preliminaries}\label{sec:prelims}

\subsection{Instantaneous quantum polynomial circuits}\label{sec:iqp}

An IQP circuit consists of commuting gates that are diagonal in the Pauli $X$ basis \cite{shepherdTemporallyUnstructuredQuantum2009}.
For a set $\mathcal{G}\subseteq\mathbb{F}_2^n\setminus\{0\}$, which we call the active set, every $g\in\mathcal{G}$ defines the generator $X^g=\bigotimes_{i=1}^nX^{g_i}$, which applies a Pauli $X$ to every qubit $i$ with $g_i=1$.
The number $|g|$ of these qubits is the body order of the gate, or its order for short, and a gate of order $k$ is a $k$-body gate.
The circuit
\begin{equation}\label{eq:circuit}
    U_\theta=\prod_{g\in\mathcal{G}}\exp(\mathrm{i}\theta_gX^g)
\end{equation}
acts on $|0\rangle^{\otimes n}$, and we measure in the computational basis.
An IQP Born machine uses the Born distribution $p_\theta(x)=|\langle x|U_\theta|0\rangle^{\otimes n}|^2$ as a generative model over bitstrings.
Without ancilla qubits, this model is not universal \cite{recio-armengolTrainClassicalDeploy2025}.
However, its expressivity plays no role in this work, as we examine which circuits we can reach rather than which distributions a circuit can represent.

The loss gradient of a candidate depends on the circuit through the expectation values
\begin{equation}\label{eq:char}
    C^a_\theta=\langle Z^a\rangle_\theta=\sum_{x\in\{0,1\}^n}p_\theta(x)(-1)^{a\cdot x}.
\end{equation}
Here $Z^a=\bigotimes_{i=1}^nZ^{a_i}$ for a frequency $a\in\mathbb{F}_2^n$, and $a\cdot x=\sum_{i=1}^na_ix_i\bmod2$.
As a function of $a$, these values form the Fourier transform of $p_\theta$.
The inversion
\begin{equation}\label{eq:inversion}
    p_\theta(x)=\frac{1}{2^n}\sum_{a\in\mathbb{F}_2^n}C^a_\theta(-1)^{a\cdot x}
\end{equation}
gives the distribution from these values.
The distribution and all its derivatives therefore depend on the circuit only through the expectation values.

A gate with $a\cdot g=0$ commutes with $Z^a$ and cancels in $C^a_\theta$, so that only the generators in
\begin{equation}\label{eq:Sa}
    S^a=\{g\in\mathcal{G}\mid a\cdot g=1\}
\end{equation}
contribute \cite{shenCharacterizingTrainabilityInstantaneous2026}.
With these generators, the expectation value reads \cite{nestSimulatingQuantumComputers2010,armengolIQPoptFastOptimization2026}
\begin{equation}\label{eq:cosine}
    C^a_\theta=\mathbb{E}_{z}\left[\cos\Bigl(\sum_{g\in S^a}2\theta_g(-1)^{g\cdot z}\Bigr)\right],
\end{equation}
where $z$ is uniform on $\{0,1\}^n$.
A classical computer estimates it efficiently by sampling $z$, which makes the classical training of Recio-Armengol et al.~\cite{recio-armengolTrainClassicalDeploy2025} possible.

Expanding the cosine and averaging over $z$ gives the equivalent formulation \cite{shenCharacterizingTrainabilityInstantaneous2026}
\begin{equation}\label{eq:char-expansion}
    C^a_\theta=\sum_{J\in\Xi^a}(-1)^{\frac{|J|}{2}}\prod_{g\in J}\sin2\theta_g\prod_{g\in S^a\setminus J}\cos2\theta_g
\end{equation}
over the set \cite{shenCharacterizingTrainabilityInstantaneous2026}
\begin{equation}\label{eq:Xi}
    \Xi^a=\Bigl\{J\subseteq S^a\Bigm|\bigoplus_{g\in J}g=0\Bigr\}.
\end{equation}
Only the subsets of $S^a$ whose generators sum to zero survive the average.
Such a subset is a relation among the generators, and we call $\Xi^a$ the relation space.

\subsection{Loss functions}\label{sec:loss}

Training a generative model compares $p_\theta$ with the true distribution $q$.
The KL divergence
\begin{equation}\label{eq:kl}
    D_{\mathrm{KL}}\left(q\,\middle\|\,p_\theta\right)=\sum_{x\in\{0,1\}^n}q(x)\log\frac{q(x)}{p_\theta(x)}
\end{equation}
compares both exactly but requires that $q$ is known.
Because it requires summing over all $2^n$ bitstrings, its evaluation is restricted to small-scale numerical experiments.

Practical problems provide only samples of $q$, and we use the MMD loss for them \cite{grettonKernelTwosampleTest2012}.
We write the squared MMD with a Gaussian kernel of bandwidth $\sigma$ as \cite{rudolphTrainabilityBarriersOpportunities2024,recio-armengolTrainClassicalDeploy2025}
\begin{equation}\label{eq:mmd}
    \operatorname{MMD}^2_\sigma(q,p_\theta)=\mathbb{E}_{a\sim\Lambda_\sigma}\left[\left(C^a_q-C^a_\theta\right)^2\right],
\end{equation}
where $C^a_q$ is the Fourier transform of $q$.
Under $\Lambda_\sigma$, every bit of $a$ equals one independently with probability $p_\sigma=\bigl(1-e^{-1/(2\sigma^2)}\bigr)/2$.
A small bandwidth therefore probes correlations of high order.

Both losses are zero exactly when $p_\theta=q$ and both depend on $\theta$ only through the values $C^a_\theta$, the MMD directly and the KL divergence through Eq.~\eqref{eq:inversion}.
Their gradients are weighted sums of the derivatives of $C^a_\theta$.
In practice, we estimate the expectation value in Eq.~\eqref{eq:mmd} from samples.
We compute $C^a_q$ directly on the data and estimate $C^a_\theta$ by Eq.~\eqref{eq:cosine}, with samples $a\sim\Lambda_\sigma$ and uniform samples $z$.

\subsection{Chow-Liu tree}\label{sec:cl}

Liu and Wang~\cite{liuDifferentiableLearningQuantum2018} use the CL tree as the entangling structure of a Born machine.
A spanning tree connects all $n$ qubits with $n-1$ edges.
Among all tree-structured distributions, the CL tree is optimal as it minimizes the KL divergence to the true distribution $q$~\cite{chowApproximatingDiscreteProbability1968}.
The tree is constructed as the maximum-weight spanning tree when every pair of bits $i,j$ is weighted by the mutual information
\begin{equation}\label{eq:mi}
    I_{ij}=\sum_{x_i,x_j\in\{0,1\}}q_{ij}(x_i,x_j)\log\frac{q_{ij}(x_i,x_j)}{q_i(x_i)\,q_j(x_j)}
\end{equation}
of the marginals $q_i$ and $q_{ij}$ of the data.

\section{Related work}\label{sec:related}

In IQP and related Born machines, the circuit structure is fixed before training.
Recio-Armengol et al.~\cite{recio-armengolTrainClassicalDeploy2025} train IQP circuits classically on up to hundreds of qubits and hundreds of thousands of parameters with all gates up to a chosen body order, for example all one- and two-qubit gates for the D-Wave dataset, and initialize two-qubit angles from the data covariance.
Sparse alternatives derive the structure from the data, along the CL tree \cite{liuDifferentiableLearningQuantum2018}, the cliques of a Markov network \cite{bakoProbleminformedGraphicalQuantum2026} or the largest empirical Fourier coefficients \cite{huangSpectralBornMachines2026}.
In all cases, the circuit does not change during training.

In adaptive methods like ADAPT-VQE, the optimization algorithm finds the structure automatically from a pool of candidates.
ADAPT-VQE adds the pool operator with the largest gradient at angle zero \cite{grimsleyAdaptiveVariationalAlgorithm2019}, and Li and Cheng~\cite{liAdaptiveCircuitLearning2023} apply this rule to Born machines with general circuits.
The pools of these methods already exclude candidates whose gradient is always zero.
In qubit-ADAPT-VQE, Pauli strings with an even number of $Y$ operators have no effect on the energy of real states, so the pool contains only odd strings \cite{tangQubitADAPTVQEAdaptiveAlgorithm2021}, and Li and Cheng~\cite{liAdaptiveCircuitLearning2023} restrict their pool to real operators for the same reason.
ADAPT-QAOA keeps only operators that commute with the $\mathbb{Z}_2$ symmetry of the cost, because all others have zero gradient in every round \cite{zhuAdaptiveQuantumApproximate2022}.
Shkolnikov et al.~\cite{shkolnikovAvoidingSymmetryRoadblocks2023} show by induction over the rounds that operators which change a symmetry shared by the Hamiltonian and the reference state are never selected, and build symmetry-adapted pools.
For IQP Born machines, Shen et al.~\cite{shenCharacterizingTrainabilityInstantaneous2026} show $\partial_{\theta_g}C^a_\theta=0$ for $g\notin S^a$ and compute gradient variances for gates already in the circuit.
Here, we trace blindness for IQP circuits to a symmetry of the active set of gates that we choose as a seed, which involves neither the loss nor the data (see Theorem~\ref{thm:stuck}).

The notion of an affine set goes back to Shepherd~\cite{shepherdBinaryMatroidsQuantum2010}, who identifies an IQP circuit with a binary matroid and expresses its amplitudes through the weight enumerator.
There the correlation for a frequency depends only on the generators not orthogonal to it, and these form an affine matroid \cite{shepherdBinaryMatroidsQuantum2010}.
These results describe the output distribution of a fixed circuit at a single common angle.
Here, affinity characterizes which gates have a vanishing gradient at the identity across all angles of the active ones.

\section{Visibility of candidate gates}\label{sec:theory}

\subsection{Candidate gradient}\label{sec:gradient}

ADAPT scores a candidate $s\in\mathbb{F}_2^n\setminus\{0\}$ with $s\notin\mathcal{G}$ by the loss gradient with respect to its angle at $\theta_s=0$.
At this angle, the gate $\exp(\mathrm{i}\theta_sX^s)$ is the identity and does not change the circuit, so all candidates are scored on the same circuit.
For the derivative of $C^a_\theta$ with respect to $\theta_s$, we extend the relation space from Eq.~\eqref{eq:Xi} to
\begin{equation}\label{eq:Xi-v}
    \Xi^a(v)=\Bigl\{J\subseteq S^a\Bigm|\bigoplus_{g\in J}g=v\Bigr\}.
\end{equation}
For a candidate $s$ at frequency $a$ with $a\cdot s=1$, every set in $\Xi^a(s)$ has odd size, because each of its generators contributes $a\cdot g=1$ to $a\cdot s$.
In this case, the derivative at $\theta_s=0$ is (Appendix~\ref{app:derivation})
\begin{equation}\label{eq:candidate-gradient}
\begin{split}
    \left.\partial_{\theta_s}C^a_\theta\right|_{\theta_s=0}={}&2\sum_{J\in \Xi^a(s)}(-1)^{\frac{|J|+1}{2}}\prod_{g\in J}\sin2\theta_g\\
    &\times\prod_{g\in S^a\setminus J}\cos 2\theta_g.
\end{split}
\end{equation}
For $a\cdot s=0$, the gate commutes with $Z^a$ and the derivative is zero.

\subsection{Visibility}\label{sec:visibility}

\begin{definition}[Blindness]\label{def:blind}
    A candidate $s\notin\mathcal{G}$ is blind for $\mathcal{G}$ if $\left.\partial_{\theta_s}C^a_\theta\right|_{\theta_s=0}=0$ for every frequency $a$ and all parameters $\theta$.
    Otherwise it is visible.
\end{definition}

A blind candidate has zero gradient under every loss that depends on $\theta$ only through the values $C^a_\theta$.
Greedy selection therefore never chooses a blind candidate, unless estimation noise gives it a nonzero score.

\begin{theorem}[Visibility]\label{thm:visibility}
    For a candidate $s\notin\mathcal{G}$ the following statements are equivalent.
    \begin{enumerate}
        \item The candidate $s$ is visible for $\mathcal{G}$.
        \item There is a frequency $a\in\mathbb{F}^n_2$ with $a\cdot s=1$ and $s\in\operatorname{span}_{\mathbb{F}_2}S^a$.
        \item There is a linearly independent set $D\subseteq\mathcal{G}$ of odd size with $\bigoplus D=s$.
    \end{enumerate}
\end{theorem}

The proof is in Appendix~\ref{app:proof-visibility}.
A candidate is visible exactly when it lies in the span of the active set $\mathcal{G}$ and can be expressed as a sum of an odd number of independent active generators.
For $\mathcal{G}=\{e_1,e_2,e_3\}$, the candidate $e_1\oplus e_2\oplus e_3$ is visible, while $e_1\oplus e_2$ is only a sum of an even number of independent active generators and is blind.

\subsection{Affinity}\label{sec:affinity}

We say that a frequency $a$ sees the generator $g$ if $a\cdot g=1$, that is if $g\in S^a$.
As we show, the circuit stays trapped when a single frequency sees every active generator (see Theorem~\ref{thm:stuck}).
Matroid theory calls sets that admit such a frequency affine \cite{shepherdBinaryMatroidsQuantum2010}.
Every linearly independent set is affine, because the linear map $a\mapsto(a\cdot g)_{g\in\mathcal{G}}$ is then surjective and some frequency maps to the all-ones vector.

\begin{definition}[Affine active set \cite{shepherdBinaryMatroidsQuantum2010}]\label{def:affine}
    An active set $\mathcal{G}$ is affine if there is a frequency $a_0\in\mathbb{F}^n_2$ with $a_0\cdot g=1$ for every $g\in\mathcal{G}$.
\end{definition}

Equivalently, $S^{a_0}=\mathcal{G}$, so one frequency sees all gates.
Geometrically, the generators lie on the affine hyperplane $\{v\in\mathbb{F}^n_2\mid a_0\cdot v=1\}$.
Every relation of an affine set has even size.
For a relation $J\subseteq\mathcal{G}$, the generators sum to zero, so $0=a_0\cdot\bigoplus_{g\in J}g=\sum_{g\in J}a_0\cdot g$ modulo two.
Each of the $|J|$ terms in this sum equals one, hence $|J|$ is even.
Conversely, a single relation of odd size shows that a set is not affine.

\begin{theorem}[Trapped]\label{thm:stuck}
    Let $\mathcal{G}$ be affine for the frequency $a_0$ and let $V(\mathcal{G})$ be the set of visible candidates for $\mathcal{G}$.
    Then
    \begin{equation}\label{eq:visible-set}
        V(\mathcal{G})=\{s\in\operatorname{span}_{\mathbb{F}_2}\mathcal{G}\setminus\mathcal{G}\mid a_0\cdot s=1\}.
    \end{equation}
    Moreover, for every $s\in V(\mathcal{G})$ the set $\mathcal{G}\cup\{s\}$ is affine for the same frequency $a_0$, has the same span and satisfies $V(\mathcal{G}\cup\{s\})=V(\mathcal{G})\setminus\{s\}$.
\end{theorem}

The proof is in Appendix~\ref{app:proof-stuck}.
An affine active set never gains a visible candidate, whichever visible candidate ADAPT adds.
The condition $a_0\cdot s=1$ in Eq.~\eqref{eq:visible-set} reflects a symmetry of the circuit.
Conjugation with $Z^{a_0}$ flips the sign of every active angle, because $a_0\cdot g=1$ for every $g\in\mathcal{G}$, and leaves the gate of a candidate $s$ with $a_0\cdot s=0$ unchanged.
The conjugated circuit has the same Born distribution, because $Z^{a_0}$ is diagonal and fixes $|0\rangle^{\otimes n}$.
Flipping the sign of all angles replaces every amplitude by its complex conjugate and leaves the distribution unchanged as well.
Both operations together flip the sign of $\theta_s$ alone, so the distribution is even in $\theta_s$ and its derivative at $\theta_s=0$ vanishes.
A visible candidate has $a_0\cdot s=1$, so adding it preserves this symmetry and the trap passes on to the next round.

\begin{theorem}[Escape]\label{thm:escape}
    Let $\mathcal{G}$ be non-affine and let $s\notin\mathcal{G}$ be a nonzero candidate with $s\in\operatorname{span}_{\mathbb{F}_2}\mathcal{G}$.
    Then $s$ is visible for $\mathcal{G}$, or there is a $t\in V(\mathcal{G})$ such that $s$ is visible for $\mathcal{G}\cup\{t\}$.
\end{theorem}

The proof is in Appendix~\ref{app:proof-escape}.
A spanning active set that is not affine has no permanently blind candidate.
From a seed that is not affine, every candidate in the span is visible at once or becomes visible after one suitable gate.

Visibility is a property of the expectation values $C^a_\theta$ as functions of the angles and does not bound the loss gradient from below.
The loss gradient of a visible candidate sums the derivatives in Eq.~\eqref{eq:candidate-gradient} over the frequencies with weights that depend on the data, so it can be small or zero at the angles that training has reached.
The gate $t$ of Theorem~\ref{thm:escape} must moreover lie in the pool and rank among the candidates that ADAPT adds in a round.
It joins the circuit at $\theta_t=0$, where the factor $\sin2\theta_t$ in Eq.~\eqref{eq:candidate-gradient} vanishes, so $s$ gains a nonzero gradient only once training has moved $\theta_t$ away from zero.

\subsection{Affine seeds}\label{sec:seeds}

The simplest seed is the layer of all one-body gates $\mathcal{G}=\{e_1,\dots,e_n\}$, which is also the starting point of Li and Cheng~\cite{liAdaptiveCircuitLearning2023}.
This set is linearly independent and therefore affine, with $a_0=1^n$.
Its span is all of $\mathbb{F}_2^n$, yet not every candidate is visible.
Since $1^n\cdot s$ equals $|s|$ modulo two, Theorem~\ref{thm:stuck} gives $V(\mathcal{G})=\{s\notin\mathcal{G}\mid |s|\text{ odd}\}$.
Every gate of even order is permanently blind, including all two-body gates.

A spanning tree with edge set $T$ gives the seed $\mathcal{G}=\{e_i\oplus e_j\mid \{i,j\}\in T\}$ of two-body gates, which is trapped as well.
A tree is bipartite, so $[n]$ splits into two classes $A$ and $B$ such that every edge has one endpoint in each class.
Letting $a_0$ be the indicator vector of $A$, one finds that $a_0\cdot (e_i\oplus e_j)=1$ for every edge, making $\mathcal{G}$ affine.
The span consists of all $s$ of even order, and $V(\mathcal{G})=\{s\notin\mathcal{G}\mid |s|\text{ even},\ a_0\cdot s=1\}$.
Only gates of even order with an odd number of qubits in $A$ are ever visible.

The union of both seeds is not affine, because for every edge $\{i,j\}\in T$, the set $\{e_i,e_j,e_i\oplus e_j\}$ is a relation of odd size.
If the edges cover every qubit, every nonzero candidate is visible at once (see Example~\ref{ex:edge-cover}).
A single edge is not enough for that (see Example~\ref{ex:one-body-edge}).

\section{Method}\label{sec:method}

\subsection{ADAPT-IQP}\label{sec:adapt}

We test the developed theory with ADAPT-IQP, our adaptation of the ADAPT framework for gradient-based architecture search to IQP circuits.
It starts from a seed, and each round first trains all angles for a fixed number of steps.
The pool holds the candidate gates, for example all gates of order two to six outside the seed.
Each round then scores every candidate in the pool by the loss gradient with respect to its angle at zero.
The $\ell$ candidates with the largest absolute score join the circuit at angle zero.
This greedy rule is the one of ADAPT-VQE, ADAPT-QAOA and adaptive Born machines \cite{grimsleyAdaptiveVariationalAlgorithm2019,tangQubitADAPTVQEAdaptiveAlgorithm2021,zhuAdaptiveQuantumApproximate2022,liAdaptiveCircuitLearning2023}.

Under the MMD loss, every score is an estimate, so a blind candidate can show a nonzero score.
We score every candidate with $R$ independent draws of the estimator and rank by the absolute mean.
Within one draw, all candidates share the same frequencies and the same samples $z$.
ADAPT-VQE and adaptive Born machines end the growth when the gradients fall below a threshold \cite{grimsleyAdaptiveVariationalAlgorithm2019,liAdaptiveCircuitLearning2023}.
We run a fixed number of rounds instead, so that all adaptive runs use the same training budget.

\subsection{Seeds}\label{sec:method-seeds}

The one-body seed $\{e_1,\dots,e_n\}$ is the usual starting point of adaptive methods \cite{liAdaptiveCircuitLearning2023} and serves as our baseline.
It is affine and sees only candidates of odd order (see Sec.~\ref{sec:seeds}).
Our approach is the CL seed, which adds the $n-1$ edges of the CL tree to the one-body layer.
The tree covers every qubit, so the CL seed is not affine and every candidate is visible from the first round (see Example~\ref{ex:edge-cover}).
Like the covariance-based initialization of Recio-Armengol et al.~\cite{recio-armengolTrainClassicalDeploy2025}, it uses pairwise statistics of the data before training starts.

\subsection{Heredity pool}\label{sec:scaling}

For $n=484$ qubits, a pool of gates of body order four alone holds $\binom{484}{4}\approx2.3\times10^9$ candidates, too many to score in every round.
Recio-Armengol et al.~\cite{recio-armengolTrainClassicalDeploy2025} suggest gates on more than two qubits to improve their results but state that the number of these gates becomes too large for high-dimensional datasets.
We restrict the pool by the heredity principle, under which a gate of order $k$ enters the pool only after a gate of order $k-1$ on a subset of its qubits has entered the circuit \cite{hamadaAnalysisDesignedExperiments1992,chipmanBayesianVariableSelection1995,bienLassoHierarchicalInteractions2013}.
The heredity pool holds the one-qubit extensions $g\oplus e_i$ of the active gates $g$ with $g_i=0$, up to body order four and without the gates already in the circuit.
It contains at most $n$ candidates per active gate.
For the one-body seed, the first pool consists of all $\binom{n}{2}$ two-body gates, and all of them are blind for this seed (see Sec.~\ref{sec:seeds}).
Gates of order three and four enter the pool only after ADAPT-IQP has added a gate that they extend.

\section{Numerical setup}\label{sec:setup}
We compare four circuits on every dataset to isolate the contributions of the initial active set, the training path and the number of parameters.
Specifically, One-body ADAPT-IQP starts from the one-body seed to test whether the theoretical trap appears in practice.
CL ADAPT-IQP starts from the non-trapped CL seed, where every qubit is covered.
Retrained CL takes the final circuit of CL ADAPT-IQP and trains it once from a fresh initialization, which separates the value of the structure from the value of the training path.
The fourth circuit is the static model of Recio-Armengol et al.~\cite{recio-armengolTrainClassicalDeploy2025}, which differs between the datasets and sets the reference for accuracy and parameter count.

\begin{figure}
\includegraphics[width=\columnwidth]{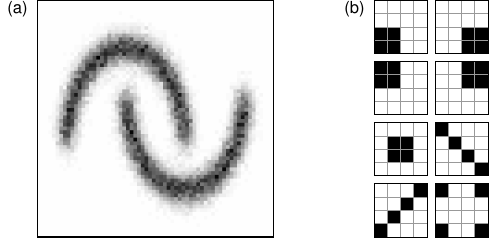}
\caption{The Two Moons and Binary Blobs datasets. (a) Empirical distribution of the 30\,000 Two Moons samples on the $64\times64$ grid, where darker cells have higher probability. (b) The eight patterns of Binary Blobs as $4\times4$ images, where a dark square marks a one. Every sample is one of the patterns with bit flips of probability $0.05$.}\label{fig:data}
\end{figure}

Two Moons consists of 30\,000 samples of the two moons distribution that we bin onto a $64\times64$ grid.
Each cell is a bitstring of length $12$ with six bits per coordinate in row-major order (Fig.~\ref{fig:data}(a)).
We train on the exact distribution with the KL divergence, so every score is exact.
The static circuit is Full IQP with all $2509$ gates of body order one to six.
The pool holds every gate of body order two to six outside the seed, so Full IQP contains every circuit that ADAPT-IQP can construct.
ADAPT-IQP adds $\ell=5$ gates per round.

Binary Blobs consists of bitstrings of length $16$ drawn from eight patterns with independent bit flips of probability $0.05$ \cite{recio-armengolTrainClassicalDeploy2025}, shown in Fig.~\ref{fig:data}(b).
It has $5000$ training and 10\,000 test samples, and we train on the samples with the MMD loss, so the scores are estimates.
The static circuit is again Full IQP, here with all 14\,892 gates of body order one to six, and the pool again holds every gate of body order two to six outside the seed.
ADAPT-IQP adds $\ell=25$ gates per round and scores every candidate with $R=10$ independent draws of the estimator with $1000$ frequencies each.

The D-Wave dataset consists of $484$-spin configurations read out from a D-Wave Advantage annealer \cite{scrivaAcceleratingEquilibriumSpinglass2023,recio-armengolTrainClassicalDeploy2025}.
The annealer minimizes a fixed Ising spin-glass Hamiltonian on a $22\times 22$ square lattice with random nearest- and next-nearest-neighbor couplings, embedded in the Pegasus topology, with an annealing time of $100\,\mu\mathrm{s}$.
Its training and test sets hold 10\,000 and 60\,000 samples, we train with the MMD loss, and the size rules out exact computation.
The static circuit is Dense IQP with all 117\,370 gates of body order one and two, the model of Recio-Armengol et al.~\cite{recio-armengolTrainClassicalDeploy2025} for this dataset.
The pool is the heredity pool of Sec.~\ref{sec:scaling} up to body order four, so the grown circuits reach gates that Dense IQP does not contain.
ADAPT-IQP adds $\ell=100$ gates per round and scores with $R=10$ draws of $4000$ frequencies, four times more than on Binary Blobs because the pool is up to seventy times larger.

On Binary Blobs and the D-Wave dataset, we take the configuration and the parameter initialization of Recio-Armengol et al.~\cite{recio-armengolTrainClassicalDeploy2025} for these datasets (Table~\ref{tab:hyper}) and change only the training length.
Their static circuits train for $1750$ steps on Binary Blobs, where a convergence criterion stops the run, and for $2500$ steps on D-Wave, while every circuit in our study trains for 12\,200 steps, the adaptive ones in $61$ rounds of $200$ steps.
Every round of an adaptive run starts from a fresh state of the Adam optimizer \cite{kingmaAdamMethodStochastic2014}, whereas a static circuit keeps one optimizer state for the whole run.
Adaptive runs train the seed in the first round, grow in each of the following $60$ rounds and end with $300$, $1500$ and $6000$ added gates.
Including the CL seed, the grown circuits have $323$, $1531$ and $6967$ parameters, about 13\,\%, 10\,\% and 6\,\% of the parameters of the static circuits, and the circuits grown from the one-body seed have $n-1$ fewer.
The loss is the mean of Eq.~\eqref{eq:mmd} over the bandwidths in Table~\ref{tab:hyper}, and every bandwidth draws its own frequencies and its own uniform samples $z$.

\begin{table}
\caption{Hyperparameters of the three datasets. The numbers of frequencies and samples hold per bandwidth.}\label{tab:hyper}
\footnotesize
\setlength{\tabcolsep}{1pt}
\begin{ruledtabular}
\begin{tabular}{lccc}
& Two Moons & Binary Blobs & D-Wave \\
\hline
Loss & exact KL & $\mathrm{MMD}^2$ & $\mathrm{MMD}^2$ \\
Bandwidths $\sigma$ & -- & $0.6$, $1.3$ & $3.9$, $6.15$, $7.8$ \\
Frequencies / samples & -- & $1000$ / $1000$ & $1000$ / $1000$ \\
Learning rate (Adam) & $10^{-3}$ & $10^{-3}$ & $10^{-3}$ \\
Training steps & $61\times200$ & $61\times200$ & $61\times200$ \\
Runs & $1$ & $10$ & $10$ \\
\hline
Static circuit body orders & $1, 2, 3, 4, 5, 6$ & $1, 2, 3, 4, 5, 6$ & $1, 2$ \\
Pool body orders & $2, 3, 4, 5, 6$ & $2, 3, 4, 5, 6$ & $2, 3, 4$ \\
Gates per round $\ell$ & $5$ & $25$ & $100$ \\
Draws $R$ $\times$ frequencies & exact & $10\times1000$ & $10\times4000$ \\
\end{tabular}
\end{ruledtabular}
\end{table}

Their initialization sets a one-body angle on qubit $i$ to $\arcsin\sqrt{\bar{x}_i}$, where $\bar{x}_i$ is the mean of bit $i$ in the training data.
A two-body angle on qubits $i$ and $j$ starts at $\pi\kappa\,c_{ij}/\max_{k\neq l}|c_{kl}|$, where $c$ is the covariance matrix of the training data in $\pm1$ encoding and the scale is $\kappa=10^{-4}$ on Binary Blobs and $10^{-3}$ on D-Wave.
All other angles start at zero, up to Gaussian noise of standard deviation $10^{-4}$ in the static circuits on Binary Blobs, and every gate that ADAPT-IQP adds starts at zero.

Two Moons is not part of their study.
There we replace the MMD estimate by the exact KL divergence and keep their initialization with the default scale of their IQPopt package \cite{armengolIQPoptFastOptimization2026}, $\kappa=0.43$, which makes the squared two-body angles of all pairs sum to $n$ on this dataset.
Nothing in this setting is random, so a single run per circuit suffices.
On the other two datasets, we report the test $\mathrm{MMD}^2$ over ten runs, estimated with $1000$ frequencies and $5000$ uniform samples $z$ per bandwidth.

\section{Numerical results}\label{sec:results}
We test the theory in an exactly simulated setting on Two Moons and in a practical setting with estimated gradients on Binary Blobs.
We then scale ADAPT-IQP to $484$ qubits on the D-Wave dataset.
Comparing the datasets also shows whether the same effect appears on different data.

\subsection{Two Moons}\label{sec:results-moons}

\begin{figure*}
\includegraphics[width=\textwidth]{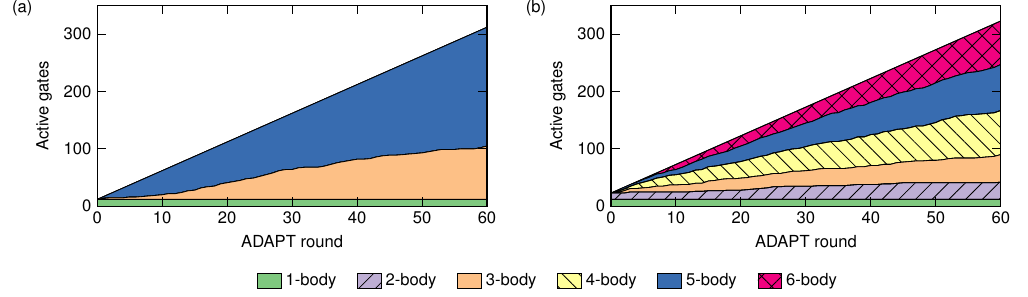}
\caption{Active gates by body order over the ADAPT rounds on Two Moons. Each area counts the active gates of one body order, and hatching marks the even orders. (a) One-body ADAPT-IQP selects odd body orders only. (b) CL ADAPT-IQP selects all orders.}\label{fig:moons-terms}
\end{figure*}

\begin{figure*}
\includegraphics[width=\textwidth]{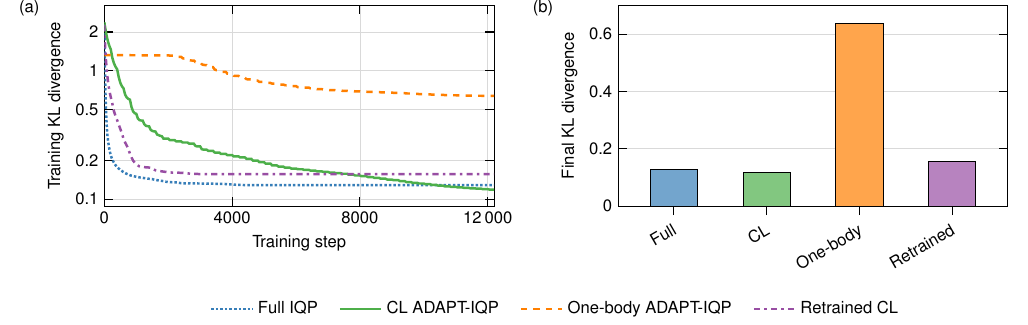}
\caption{Two Moons with the exact KL loss. (a) Training KL divergence. (b) Final training KL divergence. The curves are Full IQP (dotted), CL ADAPT-IQP (solid), One-body ADAPT-IQP (dashed) and Retrained CL (dash-dotted), and the bars labeled Full, CL, One-body and Retrained belong to the same four circuits in this order.}\label{fig:moons}
\end{figure*}

Under the exact KL loss, the one-body seed selects only gates of order three and five in all $60$ rounds (Fig.~\ref{fig:moons-terms}).
This confirms that the affine seed restricts the reachable structures, as Theorem~\ref{thm:stuck} states.
Candidates of even order reach an absolute score of at most $6\times10^{-17}$, while the largest odd score never falls below $0.025$.
As a result, the trapped circuit ends at a KL divergence of $0.64$, five times the value of CL ADAPT-IQP (Fig.~\ref{fig:moons}).

\begin{figure}
\includegraphics[width=\columnwidth]{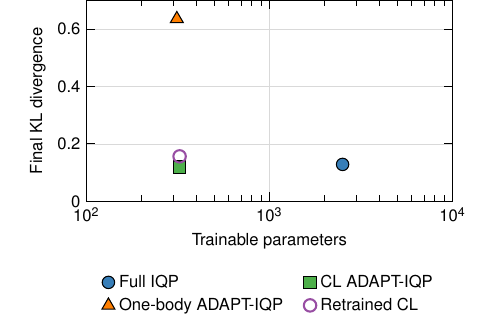}
\caption{Final KL divergence against the number of trainable parameters on Two Moons. Retrained CL is drawn as a ring, since it has the same number of parameters as CL ADAPT-IQP. Every run is deterministic, so the figure has no error bars.}\label{fig:moons-params}
\end{figure}

CL ADAPT-IQP reaches a KL divergence of $0.119$ with $323$ parameters.
By comparison, Full IQP reaches $0.129$ with $2509$ parameters (Fig.~\ref{fig:moons-params}).
Retrained CL ends at $0.157$ and stays behind the adaptive run.
Gradually growing the sparse circuit is more effective than training it from scratch on this dataset.

\subsection{Binary Blobs}\label{sec:results-blobs}

\begin{figure*}
\includegraphics[width=\textwidth]{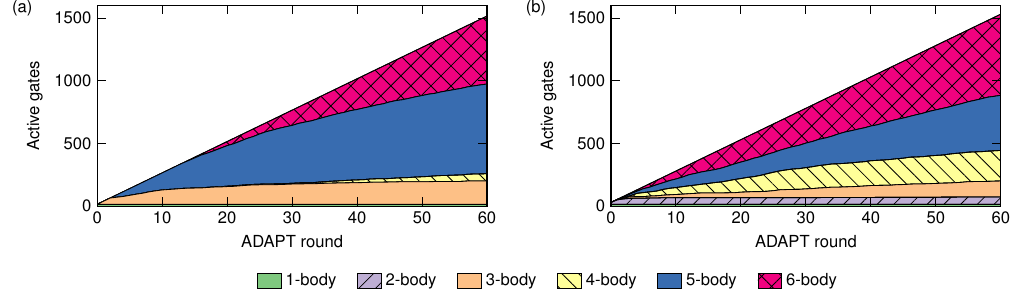}
\caption{Active gates by body order over the ADAPT rounds for the best run on Binary Blobs. Each area counts the active gates of one body order, and hatching marks the even orders. (a) One-body ADAPT-IQP selects odd orders only until noise admits an even order in round 14. (b) CL ADAPT-IQP selects all orders.}\label{fig:blobs-terms}
\end{figure*}

\begin{figure*}
\includegraphics[width=\textwidth]{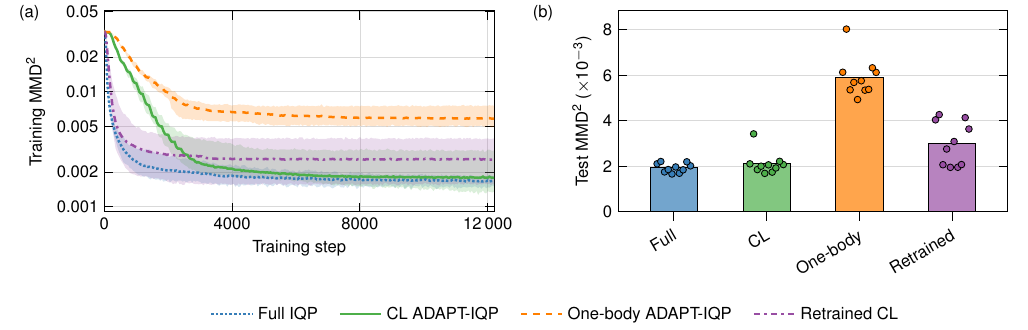}
\caption{Binary Blobs. (a) Training $\mathrm{MMD}^2$, where curves show the mean over the ten runs and bands the range over the runs. (b) Final test $\mathrm{MMD}^2$, where bars show the mean over the runs and points the single runs. The curves are Full IQP (dotted), CL ADAPT-IQP (solid), One-body ADAPT-IQP (dashed) and Retrained CL (dash-dotted), and the bars labeled Full, CL, One-body and Retrained belong to the same four circuits in this order.}\label{fig:blobs}
\end{figure*}

On Binary Blobs, the scores are estimates, which lets ADAPT-IQP pick blind candidates.
With the one-body seed, only gates of order three and five enter during the first rounds (Fig.~\ref{fig:blobs-terms}).
The first gate of even order enters between rounds $11$ and $19$, depending on the run, through estimation noise.
Each such gate makes the active set non-affine and by Theorem~\ref{thm:escape} every candidate in the span becomes reachable.
From then on, gates of even order enter as well, $337$ to $821$ of the $1500$ added gates by the end.
With the CL seed, ADAPT-IQP uses all orders from the first round.
Noise frees the one-body seed only by chance, and One-body ADAPT-IQP ends at a mean test $\mathrm{MMD}^2$ of $5.9\times10^{-3}$ against $2.1\times10^{-3}$ for CL ADAPT-IQP (Fig.~\ref{fig:blobs}).

\begin{figure}
\includegraphics[width=\columnwidth]{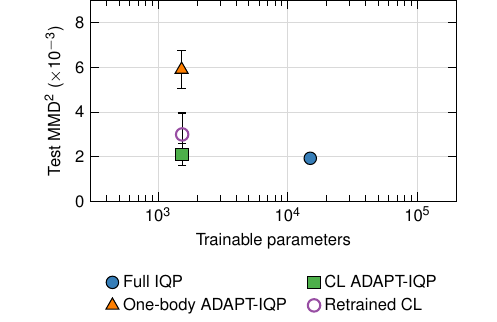}
\caption{Mean final test $\mathrm{MMD}^2$ against the mean number of trainable parameters on Binary Blobs, error bars show the standard deviation of the test $\mathrm{MMD}^2$ over the runs. Where no error bar is visible, the standard deviation is smaller than the marker. Retrained CL is drawn as a ring.}\label{fig:blobs-params}
\end{figure}

Full IQP again contains every circuit that ADAPT-IQP can construct and ends at $1.9\times 10^{-3}$.
CL ADAPT-IQP performs slightly worse with 10\,\% of the parameters (Fig.~\ref{fig:blobs-params}).
The best run of CL ADAPT-IQP ends below the mean of Full IQP.
Retrained CL ends at $3.0\times 10^{-3}$.
The structure thus explains most of the gain over the one-body seed, and the training path explains the rest.

\subsection{D-Wave}\label{sec:results-dwave}

\begin{figure*}
\includegraphics[width=\textwidth]{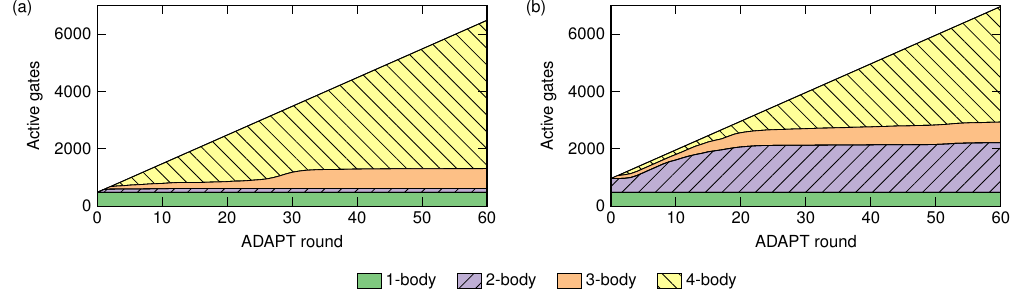}
\caption{Active gates by body order over the ADAPT rounds for the best run on the D-Wave dataset with the heredity pool capped at body order four. Each area counts the active gates of one body order, and hatching marks the even orders. (a) One-body ADAPT-IQP escapes its seed through blind two-body gates. (b) CL ADAPT-IQP starts from the CL seed.}\label{fig:dwave-terms}
\end{figure*}

On the D-Wave dataset, the scores are estimates as well, and the heredity pool determines which candidates ADAPT-IQP scores in each round.
With the one-body seed, the first pool holds only two-body gates, all of which are blind, so estimation noise alone selects the first $100$ gates (Fig.~\ref{fig:dwave-terms}).
These $100$ two-body gates cover only $127$ to $148$ of the $484$ qubits, and the pairs between the uncovered qubits stay blind.
The selected gates nevertheless make the active set non-affine, so the heredity pool frees the one-body seed in the first round.
Gates of order three enter from the second round and gates of order four from the third.

\begin{figure*}
\includegraphics[width=\textwidth]{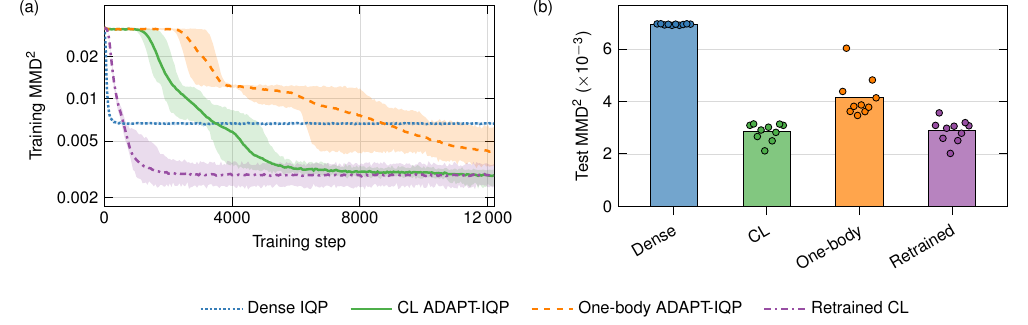}
\caption{D-Wave dataset. (a) Training $\mathrm{MMD}^2$, where curves show the mean over the ten runs and bands the range over the runs. (b) Final test $\mathrm{MMD}^2$, where bars show the mean over the runs and points the single runs. The curves are Dense IQP (dotted), CL ADAPT-IQP (solid), One-body ADAPT-IQP (dashed) and Retrained CL (dash-dotted), and the bars labeled Dense, CL, One-body and Retrained belong to the same four circuits in this order.}\label{fig:dwave}
\end{figure*}

With the CL seed, gates of order two and three enter from the first round and gates of order four from the second.
Most selected gates have body order four, 80\,\% for the one-body seed and 65\,\% for the CL seed.
With the CL seed, the training loss leaves its initial plateau in rounds $5$ to $9$, whereas with the one-body seed it does so in rounds $11$ to $16$ (Fig.~\ref{fig:dwave}).
One-body ADAPT-IQP ends at a mean test $\mathrm{MMD}^2$ of $4.2\times10^{-3}$ against $2.9\times10^{-3}$ for CL ADAPT-IQP, with a larger spread across runs.

\begin{figure}
\includegraphics[width=\columnwidth]{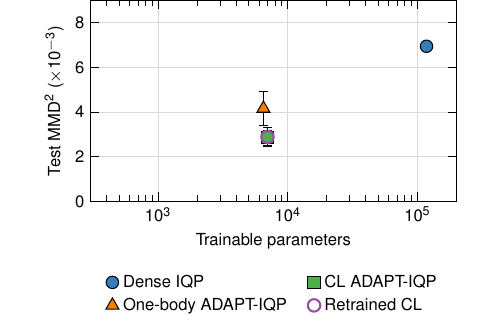}
\caption{Mean final test $\mathrm{MMD}^2$ against the mean number of trainable parameters on the D-Wave dataset, error bars show the standard deviation of the test $\mathrm{MMD}^2$ over the runs. Where no error bar is visible, the standard deviation is smaller than the marker. The ring of Retrained CL lies on the point of CL ADAPT-IQP, both circuits have $6967$ parameters and their means differ by $4\times10^{-5}$.}\label{fig:dwave-params}
\end{figure}

Dense IQP ends at $6.9\times10^{-3}$, and CL ADAPT-IQP achieves a lower test $\mathrm{MMD}^2$ with 6\,\% of its parameters (Fig.~\ref{fig:dwave-params}).
The training loss of Dense IQP stays at $6.7\times10^{-3}$ from step $2500$ to the end of the run, so the longer schedule does not put it at a disadvantage.
Dense IQP holds only one-body and two-body gates and does not contain the grown circuits, whose gates are mostly four-body.
When we set all three-body and four-body angles of the final CL ADAPT-IQP circuits to zero without further retraining, the mean test $\mathrm{MMD}^2$ rises to $3.8\times10^{-3}$.
Retrained CL ends at $2.9\times10^{-3}$ as well, so here the structure alone reproduces the result.

\subsection{Ablations}\label{sec:ablations}

\begin{figure}
\includegraphics[width=\columnwidth]{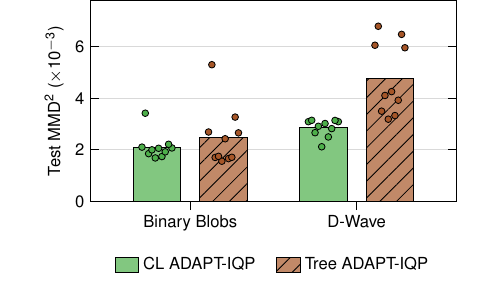}
\caption{Seed ablation. Final test $\mathrm{MMD}^2$ of ADAPT-IQP from the CL seed and from a random spanning tree (hatched) on Binary Blobs and D-Wave, bars give the mean over ten runs and points the single runs.}\label{fig:ablation-seed}
\end{figure}

We first test whether the structure of the CL tree gives an advantage over random edges.
Tree ADAPT-IQP starts from the one-body layer plus a spanning tree that we draw uniformly at random in every run and initialize like the edges of the CL tree.
The tree covers every qubit, so every candidate is visible from the first round (see Example~\ref{ex:edge-cover}).
Under the exact KL loss on Two Moons, the random tree ends at a mean of $0.116$ over ten trees against $0.119$ for the CL seed, so the results are on par regardless of structure.
On Binary Blobs, the random tree ends at $2.5\times10^{-3}$ against $5.9\times10^{-3}$ for the one-body seed, so on $12$ and $16$ qubits visibility accounts for the gap.
On D-Wave with $484$ qubits, the random tree ends at $4.8\times10^{-3}$ against $2.9\times10^{-3}$ for the CL seed (Fig.~\ref{fig:ablation-seed}).
The random tree performs no better than the one-body seed at $4.2\times10^{-3}$, whose first $100$ gates are random edges as well.
The choice of the seed therefore has a strong effect.

\begin{figure}
\includegraphics[width=\columnwidth]{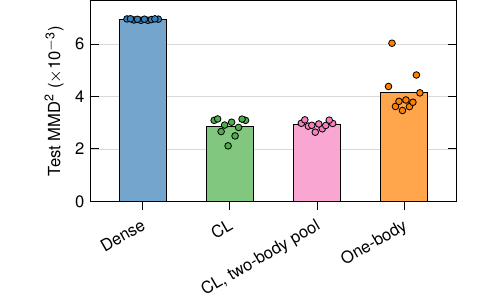}
\caption{Pool ablation on D-Wave. Final test $\mathrm{MMD}^2$ of Dense IQP, of CL ADAPT-IQP with the heredity pool capped at body order four and at body order two, and of ADAPT-IQP from the one-body seed with the pool capped at order four. Bars give the mean over ten runs and points the single runs.}\label{fig:ablation-pool}
\end{figure}

We then test whether the gain on D-Wave depends on the gates of order three and four.
We cap the heredity pool at body order two and keep the CL seed and all other settings of CL ADAPT-IQP, so every candidate is a two-body gate and every gate of the final circuit belongs to Dense IQP.
This circuit ends at a mean test $\mathrm{MMD}^2$ of $2.9\times10^{-3}$, the same value as CL ADAPT-IQP with the pool capped at order four, and both circuits have $6967$ parameters (Fig.~\ref{fig:ablation-pool}).
Dense IQP trains all 117\,370 gates at once and ends at $6.9\times10^{-3}$, so 6\,\% of its gates, selected adaptively, reach a lower loss than the full circuit.
We therefore find no advantage of the gates of order three and four on this dataset under our selection rule and training budget.
The one-body seed with the pool capped at order four ends at $4.2\times10^{-3}$, so the seed changes the result more than the body order of the pool does.
The gain over Dense IQP thus comes from the adaptive selection and the seed, not from gates outside the dense circuit.

\section{Conclusion}\label{sec:conclusion}

Adaptive algorithms grow sparse IQP circuits automatically, but they see a candidate gate only through its gradient.
A blind candidate never enters the circuit, so the active set determines which circuits the algorithm can reach.
We show that a candidate is visible exactly when it is the sum of an odd number of linearly independent active generators.
The selection rule and the expansion of the expectation values are known \cite{grimsleyAdaptiveVariationalAlgorithm2019,shenCharacterizingTrainabilityInstantaneous2026}, and our contributions are an exact criterion for gates not yet in the circuit together with the trap of affine active sets.
The common one-body seed never adds a gate of even order, while a seed with a one-body layer and two-body gates on every qubit sees every candidate from the start, which exact gradients on $12$ qubits confirm.
On Binary Blobs, the one-body seed is freed only by the estimation noise.
On D-Wave, the heredity pool of the one-body seed holds only blind two-body gates in the first round, so noise adds random edges.
In both cases, the CL seed ends with a lower loss.
A seed of the one-body layer plus random edges sees every candidate as well, yet on $484$ qubits it ends no better than the one-body seed, so the gain there comes from the edges of the CL tree.
The circuits grown from the CL seed come close to the full circuits on $12$ and $16$ qubits with about a tenth of their parameters, and they beat the dense circuit on $484$ qubits.
With the pool capped at body order two, the CL seed reaches the same loss on $484$ qubits, so the gain over the dense circuit comes from the adaptive selection of two-body gates and from the seed.
Whether a seed is trapped reduces to the solvability of a linear system over $\mathbb{F}_2$ and is decided before training starts.

Selection by first-order gradients also drives ADAPT-VQE in quantum chemistry \cite{grimsleyAdaptiveVariationalAlgorithm2019,tangQubitADAPTVQEAdaptiveAlgorithm2021}, ADAPT-QAOA in combinatorial optimization \cite{zhuAdaptiveQuantumApproximate2022} and adaptive Born machines for data loading \cite{liAdaptiveCircuitLearning2023}, and it is a common ingredient of quantum architecture search \cite{martyniukQuantumArchitectureSearch2024}.
Such selection never picks a candidate whose gradient vanishes at the identity, and our results show this obstruction for IQP circuits.
Seed and pool therefore need the same care as the loss, whose trainability prior work has studied in depth \cite{rudolphTrainabilityBarriersOpportunities2024,shenCharacterizingTrainabilityInstantaneous2026}.
The criterion involves the generators alone and applies to every model trained through expectation values of commuting circuits, including classically trained models deployed on quantum hardware \cite{recio-armengolTrainClassicalDeploy2025,rajTrainClassicalDeploy2026,herrero-gonzalezBornUltimatumConditions2025}.
Our numerical results cover three datasets with at most $484$ qubits and one training budget per dataset.
The theorems assume exact gradients, and under estimated gradients we observe their effects without proving them.
Future work could examine the circuit structures that ADAPT finds automatically.
It could also explore how higher-order gradients or gradient-free methods \cite{ostaszewskiStructureOptimizationParameterized2021,feniouGreedyGradientfreeAdaptive2025} select candidates that are blind.

\begin{acknowledgments}
  This work was supported by the LMU Sustainability Fund (EfOiE), by the German Federal Ministry of Research, Technology and Space (BMFTR) through QuCUN, QuaRDS and CAQAO, and by the Bavarian Ministry of
  Economic Affairs through the project 6GQT.
\end{acknowledgments}

\section*{Data availability}

The Binary Blobs dataset and the D-Wave dataset are those of Recio-Armengol et al.~\cite{recio-armengolTrainClassicalDeploy2025}, and the D-Wave dataset goes back to Scriva et al.~\cite{scrivaAcceleratingEquilibriumSpinglass2023}.
The Two Moons samples, the code and the results of all runs that support the findings of this article are publicly available \cite{woelckertAdaptIqpCode2026}.

\appendix

\section{Derivation of the candidate gradient}\label{app:derivation}

Let $s\notin\mathcal{G}$ be a candidate.
Let $S^a_+$ and $\Xi^a_+$ denote the sets $S^a$ and $\Xi^a$ of the enlarged active set $\mathcal{G}\cup\{s\}$, and let $J_+\in\Xi^a_+$ be a relation of the enlarged set.
For the enlarged circuit, Eq.~\eqref{eq:char-expansion} reads
\begin{equation}\label{eq:char-expansion-plus}
    C^a_\theta=\sum_{J_+\in \Xi^a_+}(-1)^{\frac{|J_+|}{2}}\prod_{g\in J_+}\sin2\theta_g\prod_{g\in S^a_+\setminus J_+}\cos 2\theta_g.
\end{equation}
We differentiate with respect to $\theta_s$ at $\theta_s=0$.
A term with $s\notin J_+$ depends on $\theta_s$ at most through $\cos2\theta_s$, whose derivative $-2\sin2\theta_s$ vanishes at zero.
A term with $s\in J_+$ contains $\sin2\theta_s$, whose derivative at zero equals $2$.
Hence
\begin{equation}\label{eq:candidate-gradient-relations}
\begin{split}
    \left.\partial_{\theta_s}C^a_\theta\right|_{\theta_s=0}={}&2\sum_{\substack{J_+\in \Xi^a_+\\s\in J_+}}(-1)^{\frac{|J_+|}{2}}\prod_{g\in J_+\setminus\{s\}}\sin2\theta_g\\
    &\times\prod_{g\in S^a_+\setminus J_+}\cos 2\theta_g.
\end{split}
\end{equation}

A set $J_+$ with $s\in J_+\subseteq S^a_+$ exists only if $a\cdot s=1$.
For $a\cdot s=0$ the derivative is therefore zero.
For $a\cdot s=1$ put $J=J_+\setminus\{s\}$.
Then $J\subseteq S^a$, $\bigoplus J=s$, $|J_+|=|J|+1$ and $S^a_+\setminus J_+=S^a\setminus J$.
The map $J_+\mapsto J$ is a bijection onto $\Xi^a(s)$, which gives Eq.~\eqref{eq:candidate-gradient}.

\section{Proofs}\label{app:proofs}

\subsection{Proof of Theorem~\ref{thm:visibility}}\label{app:proof-visibility}

\begin{proof}
    We show $(1)\Rightarrow(2)\Rightarrow(3)\Rightarrow(1)$.

    Let $s$ be visible for $\mathcal{G}$.
    Then there are a frequency $a$ and parameters $\theta$ with $\left.\partial_{\theta_s}C^a_\theta\right|_{\theta_s=0}\neq0$.
    The derivative vanishes for $a\cdot s=0$, hence $a\cdot s=1$.
    The sum in Eq.~\eqref{eq:candidate-gradient} is not empty, so there is a $J\in\Xi^a(s)$ and $s=\bigoplus J\in\operatorname{span}_{\mathbb{F}_2}S^a$.

    Let $a\cdot s=1$ and $s\in\operatorname{span}_{\mathbb{F}_2}S^a$.
    Then $\Xi^a(s)\neq\emptyset$, and we choose $D\in\Xi^a(s)$ of minimal size.
    Suppose $D$ were linearly dependent.
    Then a nonempty $D'\subseteq D$ satisfies $\bigoplus D'=0$, and $D\setminus D'$ is a smaller element of $\Xi^a(s)$, a contradiction.
    Hence $D$ is linearly independent.
    Its size is odd because $D\in\Xi^a(s)$, and $D\subseteq S^a\subseteq\mathcal{G}$.

    Let $D=\{d_1,\dots,d_m\}\subseteq\mathcal{G}$ be linearly independent with $m$ odd and $\bigoplus D=s$.
    The linear map $a\mapsto(a\cdot d_1,\dots,a\cdot d_m)$ has full rank and is onto $\mathbb{F}^m_2$.
    Hence there is a frequency $a$ with $a\cdot d_i=1$ for all $i$, so $D\subseteq S^a$.
    Since $m$ is odd, $a\cdot s=a\cdot d_1\oplus\dots\oplus a\cdot d_m=1$ and $D\in\Xi^a(s)$.

    Set $\theta_g=\pi/4$ for $g\in D$ and $\theta_g=0$ for $g\in S^a\setminus D$.
    Every $J\in\Xi^a(s)$ with $J\neq D$ contains some $g\notin D$ or misses some $g\in D$.
    In the first case the sine product contains $\sin0=0$.
    In the second case the cosine product contains $\cos(\pi/2)=0$.
    Only the term $J=D$ survives, and the derivative equals $\pm2$.
    Hence $s$ is visible for $\mathcal{G}$.
\end{proof}

\subsection{Proof of Theorem~\ref{thm:stuck}}\label{app:proof-stuck}

\begin{proof}
    Let $s$ be visible.
    By Theorem~\ref{thm:visibility} there is a set $D\subseteq\mathcal{G}$ of odd size with $\bigoplus D=s$.
    Then $s\in\operatorname{span}_{\mathbb{F}_2}\mathcal{G}$, and $a_0\cdot s$ equals $|D|$ modulo two, which is $1$.
    Conversely let $s\in\operatorname{span}_{\mathbb{F}_2}\mathcal{G}\setminus\mathcal{G}$ with $a_0\cdot s=1$.
    Since $S^{a_0}=\mathcal{G}$, the second statement of Theorem~\ref{thm:visibility} holds with $a=a_0$ and $s$ is visible.
    This proves Eq.~\eqref{eq:visible-set}.

    Let $s\in V(\mathcal{G})$.
    Then $a_0\cdot s=1$, so $a_0$ has inner product one with every element of $\mathcal{G}\cup\{s\}$ and the enlarged set is affine for $a_0$.
    The span does not change because $s\in\operatorname{span}_{\mathbb{F}_2}\mathcal{G}$.
    Applying Eq.~\eqref{eq:visible-set} to $\mathcal{G}\cup\{s\}$ gives $V(\mathcal{G}\cup\{s\})=V(\mathcal{G})\setminus\{s\}$.
\end{proof}

\subsection{Proof of Theorem~\ref{thm:escape}}\label{app:proof-escape}

\begin{proof}
    Let $W=\operatorname{span}_{\mathbb{F}_2}\mathcal{G}$.
    Choose a basis $B\subseteq\mathcal{G}$ of $W$ and a frequency $a_0$ with $a_0\cdot b=1$ for all $b\in B$, which exists as in the proof of Theorem~\ref{thm:visibility}.
    Since $B\subseteq S^{a_0}$ we have $\operatorname{span}_{\mathbb{F}_2}S^{a_0}=W$.
    By Theorem~\ref{thm:visibility} every candidate $s\in W$ with $a_0\cdot s=1$ is visible for $\mathcal{G}$.

    Let now $a_0\cdot s=0$.
    Since $\mathcal{G}$ is not affine, there is a $g\in\mathcal{G}$ with $a_0\cdot g=0$, and $s\neq g$ because $s\notin\mathcal{G}$.
    Choose $y\in B$ and put $t=s\oplus g\oplus y$.
    Then $a_0\cdot t=1$, so $t\neq0$, and $t$ lies in $\mathcal{G}$ or is visible for $\mathcal{G}$ by the first part.

    Let $\mathcal{G}'=\mathcal{G}\cup\{t\}$.
    The three elements of $D=\{g,y,t\}$ are pairwise distinct and nonzero.
    Their sum is $\bigoplus D=s\neq0$, so $D$ is linearly independent.
    Moreover $s\notin\mathcal{G}'$, since $a_0\cdot s=0$ and $a_0\cdot t=1$.
    By Theorem~\ref{thm:visibility} the candidate $s$ is visible for $\mathcal{G}'$.
    If $t\in\mathcal{G}$, then $\mathcal{G}'=\mathcal{G}$ and $s$ is visible at once.
    Otherwise $t\in V(\mathcal{G})$ and $s$ becomes visible when ADAPT adds $t$.
\end{proof}

\section{Examples of seeds that are not affine}\label{app:examples}

\begin{example}[One-body gates and one edge]\label{ex:one-body-edge}
    Let $n\geq4$ and $\mathcal{G}=\{e_1,\dots,e_n,e_1\oplus e_2\}$.
    The set $\{e_1,e_2,e_1\oplus e_2\}$ is a relation of odd size, so $\mathcal{G}$ is not affine, and $\operatorname{span}_{\mathbb{F}_2}\mathcal{G}=\mathbb{F}_2^n$.
    By Theorem~\ref{thm:escape} no candidate is permanently blind.
    Still, not every candidate is visible at once.
    Let $s$ be a candidate of even order whose support avoids the qubits $1$ and $2$, say $s=e_3\oplus e_4$, and let $D\subseteq\mathcal{G}$ be linearly independent with $\bigoplus D=s$.
    If $e_1\oplus e_2\notin D$, then $D$ consists of one-body gates only and $|D|=|s|$ is even.
    If $e_1\oplus e_2\in D$, then the one-body gates in $D$ sum to $s\oplus e_1\oplus e_2$, so $D$ contains $e_1$ and $e_2$ and is linearly dependent.
    Hence no linearly independent $D$ of odd size exists, and $s$ is blind for $\mathcal{G}$.
    The candidate $e_1\oplus e_3$ is visible through $D=\{e_1\oplus e_2,e_2,e_3\}$.
    If ADAPT adds it, $e_3\oplus e_4$ is visible through $D=\{e_1\oplus e_3,e_1,e_4\}$.
    Blindness then lasts one round, as Theorem~\ref{thm:escape} allows, while gates whose support avoids qubits $3$ and $4$ leave $e_3\oplus e_4$ blind.
\end{example}

\begin{example}[One-body gates and an edge cover]\label{ex:edge-cover}
    Let $E$ be a set of edges that covers every qubit, for instance a spanning tree, and let $\mathcal{G}=\{e_i\mid i\in[n]\}\cup\{e_i\oplus e_j\mid \{i,j\}\in E\}$.
    For every edge the set $\{e_i,e_j,e_i\oplus e_j\}$ is a relation of odd size, so $\mathcal{G}$ is not affine, and $\operatorname{span}_{\mathbb{F}_2}\mathcal{G}=\mathbb{F}_2^n$.
    Here every candidate $s\notin\mathcal{G}$ is visible at once.
    If $|s|$ is odd, then $D=\{e_i\mid i\in\operatorname{supp}s\}$ is linearly independent of odd size with $\bigoplus D=s$.
    Let $|s|$ be even.
    Choose $i\in\operatorname{supp}s$ and an edge $\{i,j\}\in E$ that contains $i$.
    Put $K=\operatorname{supp}s\,\triangle\,\{i,j\}$ and $D=\{e_i\oplus e_j\}\cup\{e_k\mid k\in K\}$.
    Then $\bigoplus D=(e_i\oplus e_j)\oplus(s\oplus e_i\oplus e_j)=s$.
    The size $|K|$ equals $|s|-2$ or $|s|$, depending on whether $j\in\operatorname{supp}s$, so it is even and $|D|=1+|K|$ is odd.
    Finally $D$ is linearly independent, because $e_i\oplus e_j$ lies in the span of $\{e_k\mid k\in K\}$ only if both $i$ and $j$ lie in $K$, and $i\notin K$ by construction.
    By Theorem~\ref{thm:visibility} the candidate $s$ is visible.
\end{example}

\bibliography{bibliography}

\end{document}